\documentclass[12pt]{article}

\ifx\pdfoutput\undefined
\usepackage[dvips,bookmarks]{hyperref}
\else
\usepackage{hyperref}
\fi
\hypersetup{colorlinks=false,bookmarksopen,bookmarksnumbered,citecolor=blue,
   pdfstartview=FitH}

\usepackage{latexsym}
\usepackage{amssymb,amsfonts,amsmath}
\usepackage{graphicx} 
\usepackage{indentfirst}
\usepackage{tensor}
\usepackage{bbm}
\usepackage{slashed, tensor}

\parskip=\medskipamount

\usepackage[mathscr]{euscript} % alternate script style

\newcommand {\cA}{{\cal A}}
\newcommand {\cB}{{\cal B}}
\newcommand {\cC}{{\cal C}}
\newcommand {\cD}{{\cal D}}
\newcommand {\cE}{{\cal E}}
\newcommand {\cF}{{\cal F}}

\newcommand {\cK}{{\cal K}}
\newcommand {\cL}{{\cal L}}
\newcommand {\cM}{{\cal M}}
\newcommand {\cN}{{\cal N}}

\newcommand {\cV}{{\cal V}}
\newcommand {\cW}{{\cal W}}

\newcommand {\cZ}{{\cal Z}}
\def\a{\alpha}
\def\b{\beta}

\def\d{\delta}

\def\g{\gamma}
\def\G{\Gamma}

\def\q{\theta}

\def\s{\sigma}

\def\D{\Delta}

\def\L{\Lambda}

\def\ri{{\rm i}}

\newcommand{\gd}{{\dot\g}}
\newcommand{\dd}{{\dot\d}}
\newcommand{\ad}{{\dot{\alpha}}}
\newcommand{\bd}{{\dot{\beta}}}

\newcommand{\1}{{\underline{1}}}
\newcommand{\2}{{\underline{2}}}

\newcommand{\ve}{\varepsilon}

\newcommand{\pa}{\partial}
\newcommand{\hf}{\frac12}

\newcommand{\be}{\begin{equation}}
\newcommand{\ee}{\end{equation}}
\newcommand{\bea}{\begin{eqnarray}}
\newcommand{\eea}{\end{eqnarray}}
\newcommand{\non}{\nonumber}
\newcommand{\ba}{\begin{array}}
\newcommand{\ea}{\end{array}}

\newcommand{\bm}[1]{\mbox{\boldmath$#1$}}

\def\double #1{#1{\hbox{\kern-2pt $#1$}}}

\newcommand{\bsubeq}{\begin{subequations}}
\newcommand{\esubeq}{\end{subequations}}

\newcommand{\N}{{\mathcal N}}
\newcommand{\rd}{\mathrm d}
\newcommand{\HC}{{\mathrm{c.c.}}}

\newcommand{\eps}{{\epsilon}}

\newcommand{\eol}{\notag \\}

\numberwithin{equation}{section}  % Resets equation number at beginning of each section

\renewcommand{\eps}{\ve}

\begin{document}
%%%%%%%%%%%%%%%%
%%%%%%%%%%%%%%%%
\begin{titlepage}
\begin{flushright}
May 2012\\
\end{flushright}

\begin{center}
{\Large \bf 
Superform formulation for vector-tensor multiplets \\
in conformal supergravity
}
\end{center}

\begin{center}

{\bf
Joseph Novak} \\
\vspace{5mm}

\footnotesize{
{\it School of Physics M013, The University of Western Australia\\
35 Stirling Highway, Crawley W.A. 6009, Australia} \\
\vspace{2mm}
{\it Email:} \texttt{joseph.novak@uwa.edu.au}
}  
~\\
\vspace{2mm}

\end{center}

\begin{abstract}
\baselineskip=14pt

The recent papers arXiv:1110.0971 and arXiv:1201.5431 have provided a superfield description for vector-tensor multiplets 
and their Chern-Simons couplings in 4D $\cN = 2$ conformal supergravity. Here we develop a superform formulation for these theories. 
Furthermore an alternative means of gauging the central charge is given, making use of a deformed vector multiplet, which may be 
thought of as a variant vector-tensor multiplet. Its Chern-Simons couplings to additional vector multiplets are also constructed. This multiplet 
together with its Chern-Simons couplings are new results not considered by de Wit {\it et al.} in hep-th/9710212.

\end{abstract}

\vfill
\end{titlepage}

\newpage
\renewcommand{\thefootnote}{\arabic{footnote}}
\setcounter{footnote}{0}

\tableofcontents{}
\vspace{1cm}
\bigskip\hrule

%%%%%%%%%%%%%%%%%%%%%%%%%%%%%%%%%%%%%%%%%%%%%%%%%%%%%%
%%%%%%%%%%%%%%%%%%%%%%%%%%%%%%%%%%%%%%%%%%%%%%%%%%%%%%

\section{Introduction}

The vector-tensor (VT) multiplet \cite{SSW, SSW2} is an off-shell representation of $\cN = 2$ supersymmetry with central 
charge, similar to the Fayet-Sohnius hypermultiplet \cite{Fayet, Sohnius}. Its physical fields consists of a real scalar, a vector and an antisymmetric 
gauge field and a doublet of Weyl spinors. The multiplet may be viewed as a dual version of the $\cN = 2$ Abelian vector multiplet, 
obtained by dualizing one of the two physical scalars of the vector multiplet into a gauge two-form. However they have different 
auxiliary fields: a real isotriplet for the vector multiplet and a real scalar for the VT multiplet. From the point of view of $\cN = 1$ 
supersymmetry, the VT multiplet consists of a vector and a tensor multiplet \cite{Milewski}.\footnote{See also the recent paper \cite{KN} for a discussion in AdS.} 
On the other hand, the $\cN  = 2$ vector multiplet decomposes into a sum of a vector and a chiral multiplet.

Interest in the VT multiplet was revived in the mid-1990s \cite{deWKLL}, when its significance in the context of string compactifications was 
realized. In rigid supersymmetry the VT multiplet and its nonlinear version discovered in \cite{Claus1, Claus2} have become the subject of various studies in 
components \cite{Claus1, Claus2} and in superspace \cite{HOW, GHH, DKT, BHO, DK, IS, DIKST}. In particular, a general harmonic 
superspace formalism for rigid supersymmetric theories with gauged central charge was developed in \cite{DIKST}.

An exhaustive analysis of VT multiplets and their Chern-Simons couplings to vector multiplets in supergravity was given in \cite{Claus3}, 
with the use of superconformal tensor calculus. As pointed out by the authors of \cite{Claus3}, a superfield formulation was desired due to the 
complicated structure of their component results. Such a superfield formulation has only appeared recently in \cite{KN, BN}, after suitable 
superspace formulations for $\cN =2$ conformal supergravity were developed \cite{KLRT-M08, Butter4D}.

In \cite{KN} the rigid supersymmetric results of \cite{DIKST} were extended to conformal supergravity within the superspace formulation of \cite{KLRT-M08}. 
This work presented two different sets of consistent constraints for the VT multiplet,
the linear and non-linear versions constituting the two inequivalent cases of \cite{Claus3}. In principle the superspace formulation 
of \cite{KLRT-M08} is sufficient to describe the VT multiplet, however it turned out that the superconformal formulation of \cite{Butter4D} has a covariant derivative algebra that is simpler to work with, 
although it possesses a more complex structure group, $\rm{SU}(2,2|2)$. It provides a direct link to the methods 
of superconformal tensor calculus used by \cite{Claus3}.  In fact, Ref. \cite{BN} 
has made use of this formulation to lift the results of \cite{Claus3} to superspace, recast them in a simpler more symmetric form and verify that the results in \cite{KN} describe the two 
inequivalent cases. However, the geometric origin of the constraints remains obscure.

In our opinion, a more geometric approach is to develop a superform description for the VT multiplet by requiring its gauge one-form and two-form
to be component projections of gauge superforms. The key difference with other superspace approaches is that it makes manifest the existence of two gauge fields 
(the one-form and the two-form), without the need to go to components. It should be mentioned that superform formulations for the VT multiplet (and its Chern-Simons couplings)
 have been explored in \cite{HOW, GHH, BHO}, where a 
complex central charge was used in flat central charge superspace to describe the original VT multiplet.\footnote{See also \cite{AGHH} for a 
description of a particular VT multiplet in supergravity with the use of ``radical constraints''.} In Ref. \cite{HOW} a geometric description of the one-form was achieved through a 
basic complex spinorial superfield, $W_\a^i$, constrained to yield the appropriate 
component structure. An alternative description in terms of 2-form geometry was given in \cite{GHH, BHO} , where the basic superfield $L$ describing the VT multiplet 
appeared explicitly in the construction and was constrained similarly. Although this approach has lead to a 
description of the basic VT multiplet in flat superspace, a superform description for the VT multiplet models of \cite{Claus3} still 
remains.\footnote{See however, \cite{ADS, ADST} for an alternative approach using Free Differential Algebra.}

This paper fills the gap in understanding of the geometric origin of the constraints found in \cite{BN}. We provide a superform formulation, differing from the formulations of 
\cite{HOW, GHH, BHO} on some points. Firstly we employ a gauge one-form and two-form interacting via Chern-Simons couplings. Secondly we use a 
real central charge, which is sufficient for all known applications. A remarkable property of this formulation is that it allows the superfield constraints of \cite{BN} to be derived entirely 
from simple superform constraints.

This paper is structured as follows. Section \ref{GCCS} provides a discussion of how to 
gauge a real central charge in $\cN = 2$ conformal supergravity and the basic superform structure that will be used in the paper. Section \ref{VTS} includes a discussion of the VT multiplet 
constraints in supergravity and gives its superform description in terms of one-form and two-form geometry, deriving the constraints for the VT multiplet from the Bianchi identities. The superform 
formulation turns out to be very powerful and in section \ref{GCCVT} it will be seen that there is a possibility of gauging the central charge with a gauge potential with non-trivial action under the 
central charge. This leads to a new one-form and two-form formulation, providing one with constraints that describe a new locally supersymmetric multiplet with Chern-Simons couplings to vector multiplets.
The simplest case reduces to the VT multiplet discovered by \cite{Theis1, Theis2} in flat superspace.

Some technical appendices are also included. Appendix \ref{conformalSpace} contains 
a summary of the conformal supergravity formulation of \cite{Butter4D}. Appendix \ref{BINote} contains a useful note on solving Bianchi identities. A brief discussion of a 
5D interpretation for central charge superspace is presented in appendix \ref{5DI}. Our notations and conventions follow those 
in \cite{Ideas} (and are summarized in \cite{BN}).

%%%%%%%%%%%%%%%%%%%%%%%%%%%%%%%%%%%%%%%%%%%%%%%%%%%%%%
%%%%%%%%%%%%%%%%%%%%%%%%%%%%%%%%%%%%%%%%%%%%%%%%%%%%%%

\section{Gauging the central charge with a vector multiplet} \label{GCCS}

In this section we review how to gauge the central charge in conformal supergravity and describe gauge $p$-forms possessing central charge 
transformations.

%%%%%%%%%%%%%%%%%%%%%%%%%%%%%%%%%%%%%%%%%%%%%%%%%%%%%%

\subsection{Setup} \label{CSCC}

We will use the superspace formulation for $\cN = 2$ conformal supergravity developed in \cite{Butter4D} as reformulated in \cite{BN} (see appendix \ref{conformalSpace}). 
To gauge the central charge we use a standard $\N = 2$ vector multiplet introduced via gauge covariant derivatives
\be \bm \nabla_A := \nabla_A + V_A \D \ ,
\ee
where $V_A$ is the gauge connection for the vector multiplet and $\D$ is a real central charge. We require that the central charge 
obey the Leibniz rule and commute with the covariant derivatives
\be [ \D , \nabla_A] = [\D, \bm \nabla_A] = 0 \ , \label{Commutes}
\ee
which requires $V_A$ to be annihilated by the central charge, $\D V_A = 0$. 
The gauge-covariant derivative algebra is then
\begin{align} [\bm \nabla_A, \bm \nabla_B\} &= T_{AB}{}^C \bm \nabla_C + F_{AB} \D + \hf R_{AB}{}^{cd} M_{cd} + R_{AB}{}^{kl} J_{kl}
	\eol & \quad
	+ \ri R_{AB}(Y) Y + R_{AB} (\mathbb{D}) \mathbb{D} + R_{AB}{}^C K_C~,
\end{align}
where the torsion and curvature remain the same as those of $\nabla_A$, and $F = \hf E^B E^A F_{AB}$ is the field-strength for the gauge connection, 
$V = E^A V_A$,
\be
F = \rd V \ , \quad F_{AB} = 2 \bm \nabla_{[A} V_{B \} } - T_{AB}{}^C V_C \ .
\ee
The presence of the one-form potential in the gauge covariant derivative algebra leads to the Bianchi identity for 
the field strength
\be
\rd F = 0 \ , \quad \bm \nabla_{[ A} F_{BC \}}- T_{[AB}{}^D F_{|D| C \} } = 0 \ .
\ee
We then impose at mass dimension-1 the standard vector multiplet constraints \cite{GSW}
\be F_\a^i{}_\b^j = - 2 \eps^{ij} \eps_{\a\b} \bar{\cZ} \ , \quad F^\ad_i{}^\bd_j = 2 \eps_{ij} \eps^{\ad\bd} \cZ \ , \quad F_\a^i{}^\bd_j = 0 \ ,
\ee
where $\cZ$ is a primary superfield with dimension 1 and $\rm U(1)$ weight $-2$,
 \be
K_A \cZ = 0 \ , \quad \mathbb{D} \cZ = \cZ \ , \quad Y \cZ = -2 \cZ \ .
\ee
The Bianchi identities may then be solved giving
\begin{subequations}
\begin{align}
F_a{}_\b^j &= \frac{\ri}{2} (\s_a)_\b{}^\gd \bar{\bm \nabla}_\gd^j \bar{\cZ} \ , \qquad 
F_a{}^\bd_j = - \frac{\ri}{2} (\s_a)_\g{}^\bd \bm \nabla^\g_j \cZ \ , \\
F_{ab} &= - \frac{1}{8} (\s_{ab})_{\a\b} ( \bm\nabla^{\a \b} \cZ + 4 W^{\a\b} \bar{\cZ})
+ \frac{1}{8} (\tilde{\s}_{ab})_{\ad\bd}  (\bar{\bm \nabla}^{\ad \bd} \bar{\cZ} + 4 \bar{W}^{\ad\bd} \cZ) \ ,
\end{align}
\end{subequations}
where $\cZ$ is a reduced chiral superfield
\be
\bar{\bm \nabla}_{\ad}^i \cZ = 0~, \quad \bm \nabla^{ij} \cZ = \bar{\bm \nabla}^{ij} \bar{\cZ} ~,
\ee
and we define
\be \bm \nabla^{ij} := \bm \nabla^{\a (i} \bm \nabla^{j)}_\a \ , \quad \bar{\bm \nabla}^{ij} := \bar{\bm \nabla}^{(i}_\ad \bar{\bm \nabla}^{j) \ad} \ .
\ee
We note the following useful identity:
\begin{align} \bm \nabla_\a^i \bm \nabla_\b^j =& \hf (\eps_{\a\b} \bm \nabla^{ij} - \eps^{ij} \bm \nabla_{\a\b}) - \eps^{ij} \eps_{\a\b} \bar{\cZ} \D 
+ \eps^{ij} \eps_{\a\b} \bar{W}_{\gd\dd} \bar{M}^{\gd \dd} \non\\
&- \frac{1}{4} \eps^{ij} \eps_{\a\b} \bar{\bm \nabla}_{\gd k} \bar{W}^{\gd\dd} \bar{S}^k_\dd 
- \frac{1}{4} \eps^{ij} \eps_{\a\b} \bm \nabla_{\g\dd} \bar{W}^\dd{}_\gd K^{\g\gd} \ .
\end{align}

In what follows we may also make use of additional abelian vector multiplets, $\cW$, which are described by a similar two-form field strength with the same 
constraints except with $\cZ$ replaced with $\cW$.

%%%%%%%%%%%%%%%%%%%%%%%%%%%%%%%%%%%%%%%%%%%%%%%%%%%%%%

\subsection{Supergravity transformations and superforms} \label{SGTS}

The supergravity transformations are realized on the gauge covariant derivatives as
\begin{align}
\d_\cK \bm \nabla_A &= [\cK, \bm \nabla_A] \ , \non\\
\cK &= \cK^C \bm \nabla_C + \hf \cK^{cd} M_{cd} + \cK^{kl} J_{kl} + \ri \cK_Y Y + \cK_{\mathbb D} \mathbb D + \cK^A K_A + C \D \ ,
\end{align}
where gauge parameters satisfy natural reality conditions. Given a primary tensor superfield $U$ (with suppressed indices), its supergravity transformation is
\be  \d_\cK U = \cK U \ .
\ee
From the local central charge transformation, 
parametrized by $C$, we can deduce the central charge transformation of $V_A$
\be \d_C \bm \nabla_A = [ C \D , \bm \nabla_A] \quad \Leftrightarrow \quad \d_C V_A = - \nabla_A C \ , \quad \D C = 0 \ . \non\\
\ee
Associated with the gauge covariant derivative $\bm \nabla_A$ is the gauge covariant exterior differential
\be \bm \nabla = \rd + V \D \ , \quad V = E^A V_A \ .
\ee

In addition to $V$ with the property $\D V = 0$, we introduce a new gauge one-form, $\cV = E^A \cV_A$ which is not annihilated by the central charge, $\D \cV \neq 0$. 
We define the transformation law of $\cV$ to be (compare with \cite{Claus3})
\be \d \cV = C \D \cV + \rd \G \ , \quad \D \G = 0 \ ,
\ee
with $\G$ the gauge parameter associated with $\cV$. The field strength
\be \cF := \bm \nabla \cV \label{oneFormGeom}
\ee
transforms covariantly
\be \d \cF = C \D \cF \ .
\ee
The one-form geometry for a vector multiplet may be thought of as a special case of the above formulation.

Similarly we may introduce a gauge two-form, $B = \hf E^B E^A B_{AB}$, and its three-form field strength, $H$, defined by
\be H := \bm \nabla B - \eta \cV \bm \nabla \cV \ , \label{twoFormGeom}
\ee
where the coupling constant, $\eta$, can be used to couple the one-form to the two-form.\footnote{The case of coupling to a number of one-forms as well as the special 
case where some of them are vector multiplets is straightforward.} We define the transformation law of $B$ to be (compare with \cite{Claus3})
\be \d B = C \D B  + \eta \G \rd \cV + \rd \L \ , \quad \D \L = 0 \ ,
\ee
where $\L$ generates the gauge transformation of $B$. The field strength, $H$, transforms covariantly
\be \d H = C \D H \ .
\ee

The Bianchi identities for the field strengths introduced are
\be
\bm \nabla \cF = F \D \cV \ , \quad \bm \nabla H = F (\D B + \eta \cV \D \cV) - \eta \cF \cF \ . \label{BIFS}
\ee
A remarkable feature of this superform structure (which reduces in components to the results of \cite{Claus3}) is that it is possible to 
rewrite it formally by treating the central charge as a covariant derivative 
with respect to an additional bosonic variable, $z$. To see this we first note that the condition \eqref{Commutes} allows us to write
\be \D \cF = \bm \nabla(\D \cV) \ , \quad \D H = \bm \nabla(\D B) - \eta \D \cV \bm \nabla \cV - \eta \cV \bm \nabla (\D \cV) \ .\label{BICentral}
\ee
Now the superform equations \eqref{BIFS} and \eqref{BICentral}, when written in terms of Lorentz indices, may be grouped together with the 
introduction of calligraphic index labels, {\rm i.e.} $\cA = (A, z)$, where $z$ is thought of as corresponding to the central charge. Making the identifications
\begin{align}
\cF_{z A}& := \D \cV_A \ , \quad H_{z AB} := \D B_{AB} + 2 \eta \cV_{[A} \D \cV_{B\} }\ , \non\\
T_{AB}{}^z &:= F_{AB} \ , \quad T_{Az}{}^B = 0 \ , \quad \bm \nabla_z := \D \ ,
\end{align}
we may extend the Bianchi identities \eqref{BIFS} and the additional equations \eqref{BICentral} to
\begin{align}
\bm \nabla_{[\cA} \cF_{\cB\cC\}} - T_{[\cA\cB}{}^\cD \cF_{|\cD|\cC\}} &= 0 \ , \non\\
 \bm \nabla_{[\cA} H_{\cB\cC\cD\}} - \frac{3}{2} T_{[\cA\cB}{}^\cE H_{|\cE|\cC\cD \}} + \frac{3}{4} \eta \cF_{[\cA\cB} \cF_{\cC\cD\}} &= 0 \ . \label{5DBI}
\end{align}
These equations appear as 5D Bianchi identities and a brief discussion of their 5D interpretation is given in appendix \ref{5DI}. This 5D form involving calligraphic indices 
and with the identification made will be used in the next section for the VT multiplet, where we refer to them as the Bianchi identities for the one-form and two-form respectively.

%%%%%%%%%%%%%%%%%%%%%%%%%%%%%%%%%%%%%%%%%%%%%%%%%%%%%%
%%%%%%%%%%%%%%%%%%%%%%%%%%%%%%%%%%%%%%%%%%%%%%%%%%%%%%

\section{Vector-tensor multiplet in supergravity} \label{VTS}

In this section we introduce the VT multiplet via its superfield. Making use of the superform formulation 
of the previous section, we will show that the superfield constraints follow from simple constraints on the one-form and 
two-form geometry.

%%%%%%%%%%%%%%%%%%%%%%%%%%%%%%%%%%%%%%%%%%%%%%%%%%%%%%

\subsection{Vector-tensor multiplet} \label{VTM}

 The general superfield construction of \cite{BN}, which agrees with the component construction of \cite{Claus1, Claus2, Claus3} is described 
 in terms of a scalar superfield, $L$. It must be coupled to the central charge vector multiplet, $\cZ$ but can also interact with 
 a number of additional vector multiplets, $\cW^I$ with $I = 2, \cdots , n$. It was noted in \cite{BN} that the superfield constraints 
on $L$ may be written in a symmetric form in terms of 
\be Y^{\hat{I}} = (L , Y^I) \ , \quad \hat{I} = 1, \cdots , n \ ,
\ee
where $Y^I$ is the imaginary part of $\cW^I / \cZ$
\be Y^I := \frac{1}{2 \ri} \Big( \frac{\cW^I}{\cZ} - \frac{\bar{\cW}^I}{\bar{\cZ}} \Big) \ .
\ee

The superfield $L$ satisfies three constraints. The first constraint may be written as
\be \bm \nabla_\a^{(i} \bar{\bm \nabla}_\bd^{j)} Y^{\hat{I}} = - \bar{\bm \nabla}_\bd^{(j} \bm \nabla_\a^{i)} Y^{\hat{I}} = 0 \ , \label{VTConstraint1}
\ee
where the constraint for $\hat{I} = 2, \cdots \ , n$ just follows from the chirality of $\cW^I$ and $\cZ$.

The second constraint is
\begin{align} \label{VTConstraint2R}
0= {\bm\nabla}^{ij} (\cZ Y^{\hat{I}}) + \bar{{\bm\nabla}}^{ij} (\bar{\cZ} Y^{\hat{I}}) 
-  Y^{\hat{I}} {\bm\nabla}^{ij} \cZ ~,
\end{align}
where the constraint for $\hat{I} = 2, \cdots \ , n$ follows from Bianchi identities satisfied by $\cZ$ and $\cW$. An interesting note made in \cite{KN, BN}, 
as a generalization of an observation in \cite{DIKST}, was that the second constraint may be motivated from the first by 
considering a consistency condition. To see this we make use of harmonic variables $u^{+i}$ and $u^-_i = \overline{u^{+i}}$ (normalized by $u^{+i}u^-_i =1$), and
demand $L$ to be independent of the harmonics
\be \bm \nabla^{--} L = 0 \ ,
\ee
where $\bm \nabla^{--} = u^{- i} \, \pa / \pa u^{+i} $  is one of the left-invariant vector fields on SU(2).
Applying successive gauged central charge covariant derivatives,
\be
\bm \nabla_\a^{\pm} := u^\pm_i \bm \nabla_\a^i~, \qquad 
\bar{\bm \nabla}_\ad^{\pm} :=  u^\pm_i \bar{\bm \nabla}_\ad^i~,
\ee 
to  the above condition and using the (anti-)commutation relations for the 
covariant derivatives gives the consistency requirement
\begin{align}
0 &= {\bm\nabla}^{\a+} {\bm\nabla}_\a^{+} \bar{{\bm\nabla}}_\ad^{+} \bar{{\bm\nabla}}^{\ad +} D^{--} L \non\\
&= D^{--} {\bm\nabla}^{\a+} {\bm\nabla}_\a^{+} \bar{{\bm\nabla}}_\ad^{+} \bar{{\bm\nabla}}^{\ad +} L 
+ 8 \ri {\bm\nabla}^{\a\ad} {\bm\nabla}_\a^{+} \bar{{\bm\nabla}}^{+}_\ad L 
	\eol & \quad
- 2 {\bm\nabla}^{\a-} {\bm\nabla}_\a^{+} \bar{{\bm\nabla}}_\ad^{+} \bar{{\bm\nabla}}^{\ad+} L
- 2 \bar{{\bm\nabla}}_\ad^{-} \bar{{\bm\nabla}}^{\ad+} {\bm\nabla}^{\a+} {\bm\nabla}_\a^{+} L \non\\
& \quad - 4 \D \Big( ({\bm\nabla}^{+})^2 (\cZ L) + (\bar{{\bm\nabla}}^{+})^2 (\bar{\cZ} L) 
 - \hf L ({\bm\nabla}^{+})^2 \cZ - \hf L (\bar{{\bm\nabla}}^{+})^2 \bar{\cZ} \Big) \ .
\end{align}
This restricts the constraints possible for $L$. Imposing the first constraint leads one to consider the 
constraint
\begin{align}
0 =& \D \Big( ({\bm\nabla}^{+})^2 (\cZ L) + (\bar{{\bm\nabla}}^{+})^2 (\bar{\cZ} L) - L ({\bm\nabla}^{+})^2 \cZ \Big) ~,
\end{align}
which is clearly satisfied by the second constraint \eqref{VTConstraint2R}.

As we will see the first two constraints are enough to guarantee the existence of a gauge one-form, they do not however imply the existence of 
a gauge two-form. The gauge two-form requires a third and final constraint. This constraint was found in \cite{BN} to be
\be \label{VTConstraint2I}
0 = \eta_{\hat{I}\hat{J}} G^{\hat{I}\hat{J} \,ij}~,
\ee
where the quantities $G^{\hat{I}\hat{J} ij}$ are defined by
\begin{align}
G^{\hat{I}\hat{J}\, ij} &:=
	\ri \cZ {\bm\nabla}^{\a(i} Y^{\hat{I}} {\bm\nabla}_\a^{j)} Y^{\hat{J}}
	+ \frac{\ri}{4} \cZ Y^{\hat{I}} {\bm\nabla}^{ij} Y^{\hat{J}} + \frac{\ri}{4} \cZ Y^{\hat{J}} {\bm\nabla}^{ij} Y^{\hat{I}}
	\eol & \quad
	+ \frac{\ri}{2} Y^{\hat{I}} {\bm\nabla}^{\a(i} Y^{\hat{J}} {\bm\nabla}_\a^{j)} \cZ
	+ \frac{\ri}{2} Y^{\hat{J}} {\bm\nabla}^{\a(i} Y^{\hat{I}} {\bm\nabla}_\a^{j)} \cZ
	+ \HC \ ,  \label{Constraint3}
\end{align}
and the numeric coefficients $\eta_{\hat{I}\hat{J}}$ are given by\footnote{This choice of $\eta_{\hat{I}\hat{J}}$ matches those of \cite{Claus3}.}
\begin{align}
\eta_{\hat{I}\hat{J}} = \left(
\begin{array}{cc}
\eta_{11} & \eta_{1I} \\
0 & \eta_{IJ}
\end{array}
\right)~, \quad \eta_{IJ} = \eta_{JI} \ . \label{eta}
\end{align}

The observation that the constraints for $L$ can be rewritten entirely in terms of $Y^{\hat{I}}$ highlights the observation made in \cite{Claus2}. That is, 
that there exists a symmetry in the constraints. Furthermore, it is possible to make redefinitions of $L$
\be L' = L + c_I Y^I \ ,
\ee
where $c_I$ is some real constant, that leave the constraints of the same form (with redefined $\eta$ parameters). In \cite{BN} this was used to show 
that the non-linear case with Chern-Simons terms 
($\eta_{1A} = 0$) may be taken as the general case.

It is worth noting that the quantities $G^{\hat{I}\hat{J} ij}$ are related to the Lagrangian for the VT multiplet \cite{BN}.\footnote{Compare to the 
linear multiplet for the vector multiplet in 5D $\cN =1$ superspace \cite{KL:5D}.} On their own they represent a 
composite linear multiplet, being primary and satisfying the condition
\be \bm \nabla_\a^{(i} G^{\hat{I}\hat{J} jk)} = 0 \ .
\ee
They correspond to a total derivative in the action. In fact, they are generated from the general VT Lagrangian in \cite{BN} via constant shifts in the real part of $\cW^I/\cZ$.

%%%%%%%%%%%%%%%%%%%%%%%%%%%%%%%%%%%%%%%%%%%%%%%%%%%%%%

\subsection{Gauge one-form}

In order to describe the one-form contained in the VT multiplet, we will make use of the one-form formulation introduced in subsection \ref{SGTS}. The field strength, $\cF$, for the 
gauge one-form, $\cV$, is constrained by the Bianchi identity
\be \bm \nabla_{[\cA} \cF_{\cB\cC\}} - T_{[\cA\cB}{}^\cD \cF_{|\cD|\cC\}} = 0 \ , \label{FBI}
\ee
with the identifications defined in subsection \ref{SGTS}. Inspired by the vector multiplet, we impose the constraints
\begin{align}
\cF_\a^i{}_\b^j = 2 \ri \eps_{\a\b} \eps^{ij} \bar{\cZ} L \ , 
\quad \cF^\ad_i{}^\bd_j = 2 \ri \eps^{\ad \bd} \eps_{ij} \cZ L \ , \quad \cF_\a^i{}^\bd_j = 0 \ , \label{Fconstraints}
\end{align}
where the central charge vector field, $\cZ$, is used to absorb the dilatation and U(1) weight of the superfield $L$, which we only assume to be real. The Bianchi identities then lead to 
the constraints on $L$,
\begin{align}
\bm \nabla_\a^{(i} \bar{\bm \nabla}_\ad^{j)}  L &= 0 \ , \non\\
\bm \nabla^{ij} (\cZ L) + \bar{\bm \nabla}^{ij} (\bar{\cZ} L) - L \bm \nabla^{ij} \cZ &= 0 \ . \label{onetwoVTconstraints}
\end{align}
The remaining components of $\cF$ are found to be
\begin{align}
\cF_z{}_\a^i &= - \ri \bm \nabla_\a^i L \ , \quad \cF_z{}^\ad_i = \ri \bar{\bm \nabla}^\ad_i L \ , \non\\
\cF_a{}_\b^j &= - \hf (\s_a)_{\b \ad} ( 2 \bar{\cZ} \bar{\bm \nabla}^{\ad j} L +  L \bar{\bm \nabla}^{\ad j} \bar{\cZ}) \ , 
\quad \cF_a{}^\bd_j = - \hf (\tilde{\s}_a)^{\bd \a} (2 \cZ \bm \nabla_{\a j} L + L \bm \nabla_{\a j} \cZ) \ , \non\\
\cF_{ab} &= - \frac{\ri}{4} (\s_{ab})^{\a\b} (\bm \nabla_{\a\b} (\cZ L) + 4 L F_{\a\b}) 
- \frac{\ri}{4} (\tilde{\s}_{ab})^{\ad\bd} ( \bar{\bm \nabla}_{\ad\bd} (\bar{\cZ} L) + 4 L \bar{F}_{\ad\bd}) \ , \non\\
\cF_{a z} &= - \frac{1}{8} (\s_a)_{\g\gd} [\bm \nabla^{\g k} , \bar{\bm \nabla}^\gd_k] L \ .
\end{align}
After having derived the components given above, the final Bianchi identities\footnote{We use the shorthand $\underline{\a} = {}_\a^i$ and $\underline{\bd} = {}^\bd_j$.}
\begin{align}
 \bm \nabla_{\underline{\a}} \cF_{az} &= 
 - \bm \nabla_a \cF_{z \underline{\a}} - \D \cF_{\underline{\a} a} + T_{\underline{\a} a}{}^{\underline{\bd}} \cF_{\underline{\bd} z} \ , \non\\
\bm \nabla_{\underline{\g}} \cF_{ab} &= 
- 2 \bm \nabla_{[a} \cF_{b] \underline{\g}} + T_{ab}{}^{\underline{\b}} \cF_{\underline{\b} \underline{\g}} + F_{ab} \cF_{z \underline{\g}} 
+ 2 T_{\underline{\g}[a}{}^{\underline{\gd}} \cF_{\underline{\gd} b]} + 2 F_{\underline{\g}[a} \cF_{z b]} \ , \non\\
\D \cF_{ab} &= 
- 2 \bm \nabla_{[a} \cF_{b] z} + T_{ab}{}^{\underline{\g}} \cF_{\underline{\g}z} + T_{ab}{}^{\underline{\gd}} \cF_{\underline{\gd}z} \ , \non\\
\bm \nabla_{[a} \cF_{bc]} &= T_{[ab}{}^{\underline{\g}} \cF_{\underline{\g} c]} + T_{[ab}{}^{\underline{\gd}} \cF_{\underline{\gd} c]} - F_{[ab} \cF_{c] z} \ ,
\end{align}
can be seen to be identically satisfied by making use of the following consequences of the constraints \eqref{onetwoVTconstraints} (see also appendix \ref{BINote}):
\begin{align}
 \bm \nabla_\a^i \bar{\bm \nabla}_\ad^j L &= - \frac{1}{4} \eps^{ij} [\bm \nabla_\a^k, \bar{\bm \nabla}_{\ad k}] L + \ri \eps^{ij} \bm \nabla_{\a\ad} L \ , \non\\
 \bar{\bm \nabla}_{\bd k} \cF_{\a\b} =& \ri \bar{\bm \nabla}_{\bd k} L F_{\a\b} - \hf \bm \nabla_{(\a \bd} (2 \cZ \bm \nabla_{\b) k} L 
 + L \bm \nabla_{\b) k} \cZ ) + \frac{\ri}{2} \bm \nabla_{(\a k} \cZ \cF_{\b) \bd , z} \non\\
&+ \frac{\ri}{2} W_{\a\b} (2 \bar{\cZ} \bar{\bm \nabla}_{\bd k} L + L \bar{\bm \nabla}_{\bd k} \bar{\cZ}) \ , \non\\
\bm \nabla_{\a}^i \bm \nabla_{\b}^k \bar{\bm \nabla}_{\bd k} L &= - 4 \eps_{\a\b} \bar{\cZ} \D \bar{\bm \nabla}_\bd^i L - 4 \eps_{\a\b} \bar{W}_{\ad\bd} \bar{\bm \nabla}^{\ad i} L \ .
\end{align}

We have constructed a one-form that is expressed in terms of the basic superfield $L$. The geometry uncovered clearly highlights 
the relationship of the first constraint \eqref{VTConstraint1} and second constraint \eqref{VTConstraint2R} for the VT multiplet with the one-form in the theory, ensuring its existence.

%%%%%%%%%%%%%%%%%%%%%%%%%%%%%%%%%%%%%%%%%%%%%%%%%%%%%%

\subsection{Gauge two-form}

To describe a general VT multiplet coupled to several vector multiplets, $\cW^I$, we make use of a straightforward generalization of the two-form construction 
presented in subsection \ref{SGTS}, by coupling corresponding gauge one forms, $V^I$, to the two-form, $B$. Denoting the set of one-forms by $\cV^{\hat{I}} = (\cV, V^I)$ 
and their corresponding field strengths by $\cF^{\hat{I}} = (\cF, F^I)$, the three-form field strength is now defined by\footnote{This definition reduces in components to those of \cite{Claus3}. 
Its generalization to a number of VT multiplets is straightforward.}
\be H = \bm \nabla B - \hf \eta_{\hat{I}\hat{J}} \cV^{\hat{I}} \rd \cV^{\hat{J}} \ ,
\ee
with $\eta_{\hat{I}\hat{J}}$ given by \eqref{eta}. It follows that $H$ obeys the Bianchi identity
\be
\quad \bm \nabla_{[\cA} H_{\cB\cC\cD\}} - \frac{3}{2} T_{[\cA\cB}{}^\cE H_{|\cE|\cC\cD \}} + \frac{3}{4} \eta_{\hat{I}\hat{J}} \cF^{\hat{I}}_{[\cA\cB} \cF^{\hat{J}}_{\cC\cD\}} = 0 \ , \label{HBI}
\ee
with the identifications discussed in \ref{SGTS} and with $F^I{}_{zA} = \D V^I{}_A= 0$. 
We then impose the constraints
\begin{align} H_{\underline{\a} \underline{\b} \underline{\g}} &= H_{\underline{\ad} \underline{\bd} \underline{\gd}} 
= H_{\underline{\a} \underline{\b} \underline{\gd}} = H_{\underline{\a} \underline{\bd} \underline{\gd}} =  0 \ , \non\\
H_{a \underline{\b} \underline{\g}} &= H_{a \underline{\bd} \underline{\gd}} = 0 \ , \quad H_a{}_\b^j{}^\gd_k = - 2 \ri \d^j_k (\s_a)_\b{}^\gd \tilde{H} \ ,
\end{align}
where $\tilde{H}$ is only assumed to be a real superfield. The Bianchi identities lead to an additional constraint on $L$
\be \eta_{\hat{I}\hat{J}} G^{\hat{I}\hat{J} ij} = 0 \ ,
\ee
and constrain the lower mass-dimension components to be
\begin{align}
H_z{}_\a^i{}_\b^j &= \eps_{\a \b} \eps^{ij} \bar{\cZ} \Big(\eta_{11} L^2 + \ri \eta_{1I} \frac{\bar{\cW}^I}{\bar{\cZ}} L 
- \eta_{IJ} \frac{\bar{\cW}^I \bar{\cW}^J}{\bar{\cZ}^2}\Big) \equiv \eps_{\a\b} \eps^{ij} \bar{\cZ} H \ , \non\\
H_z{}^\ad_i{}^\bd_j &= - \eps^{\ad \bd} \eps_{ij} \cZ \bar{H} \ , \quad H_{z \underline{\a} \underline{\bd}} = 0 \ ,
\end{align}
together with
\be \tilde{H} = - \hf \eta_{\hat{I}\hat{J}} \cZ \bar{\cZ} Y^{\hat{I}} Y^{\hat{J}} \ .
\ee
The higher mass-dimension components are more complex
\begin{align}
H_{a z}{}_\b^j &= - \frac{\ri}{4} (\s_a)_{\b \ad} \Big(4 \eta_{11} \bar{\cZ} L \bar{\bm \nabla}^{\ad j} L + \eta_{11} L^2 \bar{\bm \nabla}^{\ad j} \bar{\cZ}
- \eta_{IJ} \bar{\bm \nabla}^{\ad j} \Big(\frac{\bar{\cW}^I \bar{\cW}^J}{\bar{\cZ}}\Big) \non\\
&\quad+ 2 \ri \eta_{1I} \bar{\bm \nabla}^{\ad j} L \bar{\cW}^I 
+ \ri  \eta_{1I} L \bar{\bm \nabla}^{\ad j} \bar{\cW}^I \Big) \ , \non\\
H_{ab}{}_\g^k &= 2 (\s_{ab})_\g{}^\a \bm \nabla_\a^k \tilde{H} \ , \non\\
H_{ab z} &= (\s_{ab})^{\a\b} \Big( \frac{1}{4} \eta_{11} \bm \nabla_{\a\b} (\cZ L^2) - \frac{1}{4} \eta_{11} L \bm \nabla_{\a\b} (\cZ L) - \hf L^2 \eta_{11} F_{\a\b} 
- \frac{\ri}{8} \eta_{1I} \bm \nabla_{\a\b} (L \cW^I) \non\\
&\quad- \frac{\ri}{2} \eta_{1I} L F^I_{\a\b} - \frac{1}{16} \eta_{IJ} \bm \nabla_{\a\b} \Big(\frac{\cW^I \cW^J}{\cZ}\Big) 
- \frac{1}{4} \eta_{IJ} W_{\a\b} \Big(\frac{\bar{\cW}^I \bar{\cW}^J}{\bar{\cZ}}\Big) \Big) + \HC \ , \non\\
H_{abc} &= \frac{1}{8} \eps_{abcd} (\s^d)^{\a \ad} \eta_{11} \big(\cZ \bar{\cZ} L [\bm \nabla_\a^k , \bar{\bm \nabla}_{\ad k}] L
+  2 \bm \nabla_\a^k \cZ \bar{\bm \nabla}_{\ad k} \bar{\cZ} L^2 - 4 \cZ L \bar{\bm \nabla}_{\ad k} \bar{\cZ} \bm \nabla_\a^k L \non\\
&\quad+ 4 \bar{\cZ} L \bm \nabla_\a^k \cZ \bar{\bm \nabla}_{\ad k} L + 4 \cZ \bar{\cZ} \bm \nabla_\a^k L \bar{\bm \nabla}_{\ad k} L  \big) 
- \frac{\ri}{2} \eps_{abcd} \eta_{11} \big( \bar{\cZ} L^2 \bm \nabla^d \cZ - \cZ L^2 \bm \nabla^d \bar{\cZ} \big) \non\\
&\quad- \frac{1}{8} \eps_{abcd} (\s^d)^{\a\ad} \eta_{1I} \big( \frac{\ri}{4} (\bar{\cZ} \cW^I - \cZ \bar{\cW}^I) [\bm \nabla_\a^k, \bar{\bm \nabla}_{\ad k}] L 
- \ri \bm \nabla_\a^k \cZ \bar{\bm \nabla}_{\ad k} L \bar{\cW}^I \non\\
&\quad - \ri \bar{\bm \nabla}_{\ad k} \bar{\cZ} \bm \nabla_\a^k L \cW^I - \ri \bar{\bm \nabla}_{\ad k} (\bar{\cZ} L) \bm \nabla_\a^k \cW^I 
- \ri \bm \nabla_\a^k (\cZ L) \bar{\bm \nabla}_{\ad k} \bar{\cW}^I \big) \non\\
&\quad+ \frac{1}{4} \eps_{abcd} \eta_{1I} \big( L \bm \nabla^d \bar{\cZ} \cW^I - L \cZ \bm \nabla^d \bar{\cW}^I + L \bm \nabla^d \cZ \bar{\cW}^I 
- L \bar{\cZ} \bm \nabla^d \cW^I \big) \non\\
&\quad+ \frac{1}{16} \eps_{abcd} (\s^d)^{\a\ad} \eta_{IJ} \Big( \bar{\bm \nabla}_{\ad k} \bar{\cZ} \bm \nabla_\a^k \Big(\frac{\cW^I \cW^J}{\cZ}\Big) 
- \bm \nabla_\a^k \cZ \bar{\bm \nabla}_{\ad k} \Big(\frac{\bar{\cW^I} \bar{\cW^J}}{\bar{\cZ}}\Big) \non\\
&\quad+ 2 \bm \nabla_\a^k \cW^I \bar{\bm \nabla}_{\ad k} \bar{\cW}^J \Big)\non\\
&\quad- \frac{\ri}{8} \eps_{abcd} \eta_{IJ} \Big( \bm \nabla^d \bar{\cZ} \Big(\frac{\cW^I \cW^J}{\cZ}\Big) - \bar{\cZ} \bm \nabla^d \Big(\frac{\cW^I \cW^J}{\cZ}\Big)  
+ \cZ \bm \nabla^d \Big(\frac{\bar{\cW}^I \bar{\cW}^J}{\bar{\cZ}}\Big) \non\\
&\quad- \bm \nabla^d \cZ \Big(\frac{\bar{\cW}^I \bar{\cW}^J}{\bar{\cZ}}\Big) - 2 \cW^I \bm \nabla^d \bar{\cW}^J + 2 \bm \nabla^d \cW^I \bar{\cW}^J \Big) \ .
\end{align}
The last few Bianchi identities
\begin{align}
2 \D H_{ab \underline{\g}} &= 2 \bm \nabla_{\underline{\g}} H_{ab z} - 4 \bm \nabla_{[a} H_{b] z \underline{\g}}  - 2 T_{ab}{}^{\underline{\d}} H_{z \underline{\g} \underline{\d}} 
- 4 T_{\underline{\g} [a}{}^{\underline{\dd}} H_{b] z \underline{\dd}} \non\\
&\quad- \eta_{\hat{I}\hat{J}} \cF^{\hat{I}}_{ab} \cF^{\hat{J}}_{z \underline{\g}} - \eta_{\hat{I}\hat{J}} \cF^{\hat{I}}_{z \underline{\g}} \cF^{\hat{J}}_{ab} - 2 \eta_{\hat{I}\hat{J}} \cF^{\hat{I}}_{z [a} \cF^{\hat{J}}_{b] \underline{\g}} 
+ 2 \eta_{\hat{I}\hat{J}} \cF^{\hat{I}}_{\underline{\g} [a} \cF^{\hat{J}}_{b] z} \ , \non\\
2 \bm \nabla_{\underline{\g}} H_{abc} &= 6 \bm \nabla_{[a} H_{bc] \underline{\g}} + 6 F_{[ab} H_{c] z \underline{\g}}
+ 6 T_{[ab}{}^{\underline{\dd}} H_{c] \underline{\g} \underline{\dd}} + 6 T_{\underline{\g} [a}{}^{\underline{\dd}} H_{bc] \underline{\dd}} \non\\
&\quad+ 3 \eta_{\hat{I}\hat{J}} \cF^{\hat{I}}_{[ab} \cF^{\hat{J}}_{c] \underline{\g}} - 3 \eta_{\hat{I}\hat{J}} \cF^{\hat{I}}_{\underline{\g} [a} \cF^{\hat{J}}_{bc]} \ , \non\\
2 \D H_{abc} &= 6 \bm \nabla_{[a} H_{bc] z} + 6 T_{[ab}{}^{\underline{\g}} H_{c] z \underline{\g}} + 6 T_{[ab}{}^{\underline{\gd}} H_{c] z \underline{\gd}} 
+ 3 \eta_{\hat{I}\hat{J}} \cF^{\hat{I}}_{[ab} \cF^{\hat{J}}_{c]z} - 3 \eta_{\hat{I}\hat{J}} \cF^{\hat{I}}_{z [a} \cF^{\hat{J}}_{bc]} \ , \non\\
\bm \nabla_{[a} H_{bcd]} &= \frac{3}{2} T_{[ab}{}^{\underline{\g}} H_{c d] \underline{\g}} + \frac{3}{2} T_{[ab}{}^{\underline{\gd}} H_{c d] \underline{\gd}} 
+ \frac{3}{2} F_{[ab} H_{c d] z} - \frac{3}{4} \eta_{\hat{I}\hat{J}} \cF^{\hat{I}}_{[ab} \cF^{\hat{J}}_{cd]} \ ,
\end{align}
 can also be seen to be identically satisfied.
 
The one-form and two-form geometry discussed in this section is a main result of this paper. Remarkably the complex component expressions of \cite{Claus3}, 
described by the constraints on $L$, may be recovered entirely from a simple looking superform structure and constraints on field strengths. Moreover, the precise 
relationship of the constraints for $L$ with its form geometry is seen; the first and second constraints, \eqref{VTConstraint1} and \eqref{VTConstraint2R}, imply the existence 
of a one-form and the addition of a third constraint \eqref{Constraint3} is necessary for the two-form.

%%%%%%%%%%%%%%%%%%%%%%%%%%%%%%%%%%%%%%%%%%%%%%%%%%%%%%
%%%%%%%%%%%%%%%%%%%%%%%%%%%%%%%%%%%%%%%%%%%%%%%%%%%%%%

\section{Gauging the central charge with a variant vector-tensor multiplet} \label{GCCVT}

So far a standard vector multiplet, with the property $\D V_A = 0$, has been used to gauge the central charge.
There is an alternative approach first developed in flat superspace \cite{Theis1, Theis2}, which not only gauges the central charge but can 
be used to describe a ``new'' non-linear VT multiplet, which we will refer to as the variant VT multiplet.

%%%%%%%%%%%%%%%%%%%%%%%%%%%%%%%%%%%%%%%%%%%%%%%%%%%%%%

\subsection{Gauge one-form}

To gauge the central charge, 
we introduce a new one-form, $\cV = E^A \cV_A$, which is not annihilated by the central charge. We use it to define the covariant  derivatives
\be \bm \nabla_A := \nabla_A + \cV_A \D \ .
\ee
Their (anti-)commutation relations are
\begin{align} [\bm \nabla_A, \bm \nabla_B\} &= T_{AB}{}^C \bm \nabla_C + \hf R_{AB}{}^{cd} M_{cd} + R_{AB}{}^{kl} J_{kl}
	\eol & \quad
	+ \ri R_{AB}(Y) Y + R_{AB} (\mathbb{D}) \mathbb{D} + R_{AB}{}^C K_C + \cF_{AB} \D ~,
\end{align}
where the torsion and curvature remain the same, and
\be \cF = \bm \nabla \cV \ , \quad \cF_{AB}  = 2 \bm \nabla_{[A} \cV_{B\}} - T_{AB}{}^C \cV_C \ .
\ee
The field strength can then be shown to satisfy the Bianchi identity
\be \bm \nabla \cF = \cF \D \cV \ .
\ee
As in subsection \ref{SGTS}, the Bianchi identity may be written in a 5D way with the use of calligraphic indices
\be
\quad \bm \nabla_{[\cA} \cF_{\cB\cC\}} - T_{[\cA\cB}{}^\cD \cF_{|\cD|\cC\}} = 0 \ ,
\ee
where now we define
\be \cF_{z A} := T_{z A}{}^z := \D \cV_A \ , \quad T_{AB}{}^z := \cF_{AB} \ , \quad T_{zA}{}^B = 0 \ , \quad \bm \nabla_z := \D \ .
\ee
Note that since we want the gauge potential to not be annihilated by the central charge, $\cF_{z A}$ is non-zero. Furthermore, as a 
consequence the central charge does not commute with $\bm \nabla_A$
\be [\D , \bm \nabla_A] = \cF_{z A} \D \ .
\ee
We may then impose constraints on the field strength and analyze the consequences of the Bianchi identities. Inspired by the constraints for the vector multiplet 
we choose the ``natural" constraints on the field strength
\be
\cF_\a^i{}_\b^j = 2 \ri \eps_{\a\b} \eps^{ij} \bar{M} \ , \quad \cF^\ad_i{}^\bd_j = 2 \ri \eps^{\ad\bd} \eps_{ij} M \ , \quad \cF_\a^i{}^\bd_j = 0 \ ,
\ee
where $M$ is not annihilated by the central charge. The Bianchi identities then lead to constraints on $M$
\begin{align}
\bm \nabla_\a^{(i} \bar{\bm \nabla}_\ad^{j)} {\rm ln} \Big( \frac{M}{\bar{M}} \Big) &= 0 \ , \non\\
\bar{M} \bm \nabla^{ij} \Big( \frac{M}{\bar{M}} \Big) + M \bar{\bm \nabla}^{ij} \Big( \frac{\bar{M}}{M}\Big) &= 0 \ , \label{Mconstraints}
\end{align}
and the remaining components
\begin{align} \cF_{z}{}_\a^i &= \frac{1}{\bar{M}} \bm \nabla_\a^i \bar{M}  = \bm \nabla_\a^i {\rm ln} (\bar{M}) \ , 
\quad \cF_{z}{}^\ad_i = \bar{\bm \nabla}^\ad_i {\rm ln} (M) \ , \non\\
\cF_a{}_\b^j &= \hf (\s_a)_\b{}^\ad \bar{M} \bar{\bm \nabla}_\ad^j {\rm ln} \Big(\frac{\bar{M}}{M}\Big) \ ,
\quad \cF_a{}^\bd_j = \hf (\s_a)_\a{}^\bd M \bm \nabla_j^\a {\rm ln} \Big(\frac{M}{\bar{M}}\Big) \ , \non\\
\cF_{az} &= \frac{i}{8} (\s_a)_{\g\gd} (\bm \nabla^{\g k} \bar{\bm \nabla}^\gd_k {\rm ln} (M) + \bar{\bm \nabla}^\gd_k \bm \nabla^{\g k} {\rm ln}(\bar{M})) \ , \non\\
\cF_{ab} &= - \frac{\ri}{8} (\s_{ab})^{\a\b} \bar{M} \Big(\bm \nabla_{\a\b} \Big(\frac{M}{\bar{M}}\Big) - 4 W_{\a\b}\Big) + \HC
\end{align}
The components satisfy the Bianchi identities
\begin{align}
\D \cF_{\underline{\a} a} &= - \bm \nabla_a \cF_{z \underline{\a}} - \bm \nabla_{\underline{\a}} \cF_{az} + T_{\underline{\a} a}{}^{\underline{\bd}} \cF_{\underline{\bd} z} \ , \non\\
\bm \nabla_{\underline{\g}} \cF_{ab} &= - 2 \bm \nabla_{[a} \cF_{b] \underline{\g}} + T_{ab}{}^{\underline{\b}} \cF_{\underline{\b} \underline{\g}} + \cF_{ab} \cF_{z \underline{\g}} 
+ 2 T_{\underline{\g}[a}{}^{\underline{\gd}} \cF_{\underline{\gd} b]} - 2 \cF_{\underline{\g}[a} \cF_{b] z} \ , \non\\
\D \cF_{ab} &= - 2 \bm \nabla_{[a} \cF_{b] z} + T_{ab}{}^{\underline{\g}} \cF_{\underline{\g}z} + T_{ab}{}^{\underline{\gd}} \cF_{\underline{\gd}z} \ , \non\\
\bm \nabla_{[a} \cF_{bc]} &= T_{[ab}{}^{\underline{\g}} \cF_{\underline{\g} c]} + T_{[ab}{}^{\underline{\gd}} \cF_{\underline{\gd} c]} - \cF_{[ab} \cF_{c] z} \ .
\end{align}

We note that although the central charge does not commute with the gauge covariant derivatives due to $\cF_z{}_\a^i$, the operator $\bar{M} \D$ commutes 
with $\bm \nabla_\a^i$,
\be [\bar{M} \D, \bm \nabla_\a^i] = 0 \ .
\ee

The superfield $M$ contains too many degrees of freedom. In order to describe a theory of multiplets with $8+8$ degrees of freedom (like the VT multiplet) we must fix some of them. 
From now on we constrain the theory by making the choice\footnote{We note that the transformation $M \rightarrow R M$ for $R$ an arbitrary real superfield preserves the form of the 
constraints \eqref{Mconstraints}. However, the one-form and its field strength undergoes non-trivial deformations.}
\be \bar{M} = \bar{\cZ} e^{\ri L} \ .
\ee
In the rigid supersymmetric limit with $\cZ = 1$ this corresponds to the choice made in \cite{Theis2}. However it does not completely determine the 
multiplet and additional constraints will be imposed with the use of the one-form and a two-form geometry. So far the components of $\cF$ are
\begin{align}
\cF_{z}{}_\a^i &= \ri \bm \nabla_\a^i L \ , \quad \cF_z{}^\ad_i = - \ri \bar{\bm \nabla}^\ad_i L \ , \non\\
\cF_a{}^j_\b &= \hf e^{- \ri L} (\s_a)_\b{}^\ad \bar{\bm \nabla}^j_\ad (\bar{\cZ} e^{2 \ri L}) \ , \quad \cF_a{}^\bd_j = \hf e^{\ri L} (\s_a)_\a{}^\bd \bm \nabla_j^\a (\cZ e^{- 2 \ri L}) \ , \non\\
\cF_{az} &= \frac{1}{8} (\s_a)_{\g\gd} [\bm \nabla^{\g k}, \bar{\bm \nabla}^\gd_k] L \ , \non\\
\cF_{ab} &= - \frac{\ri}{8} (\s_{ab})^{\a\b} e^{\ri L} (\bm \nabla_{\a\b} (\cZ e^{-2 \ri L}) - 4 W_{\a\b} \bar{\cZ} ) + \HC \ ,
\end{align}
where the superfield $L$ is constrained by
\begin{align}
\bm \nabla_\a^{(i} \bar{\bm \nabla}_\ad^{j)}  L &= 0 \ , \label{CCVT1} \\
e^{\ri L}\bm \nabla^{ij} (\cZ e^{-2 \ri L}) + e^{-\ri L} \bar{\bm \nabla}^{ij} (\bar{\cZ} e^{2 \ri L}) &= 0 \label{CCVT2} \ .
\end{align}
The last few Bianchi identities may be shown to be identically satisfied with the help of the identities
\begin{align}
\bm \nabla_{\a}^i \bm \nabla_{\b}^k \bar{\bm \nabla}_{\bd k} L &= 4 \ri \eps_{\a\b} \bar{\cZ} e^{\ri L} \D \bar{\bm \nabla}_\bd^i L - 4 \eps_{\a\b} \bar{W}_{\ad\bd} \bar{\bm \nabla}^{\ad i} L \ , \non\\
\bm \nabla_\g^k \big( e^{- \ri L} \bar{\bm \nabla}_{\ad\bd} (\bar{\cZ} e^{2 \ri L})\big) &= - e^{-\ri L} \bar{\bm \nabla}_{\ad\bd} (\bar{\cZ} e^{2 \ri L}) \cF_z{}_\g^k 
- 4 \ri \bm \nabla_{\g(\ad} \big( e^{- \ri L} \bar{\bm \nabla}^k_{\bd)} (\bar{\cZ} e^{2 \ri L}) \big) \non\\
&\quad- \ri e^{-\ri L} \bar{\bm \nabla}^k_{(\ad} (\bar{\cZ} e^{2 \ri L}) [\bm \nabla_\g^l , \bar{\bm \nabla}_{\bd ) l}] L \ , \non\\
\bm \nabla_\g^k \big( e^{\ri L} \bm \nabla_{\a\b} (\cZ e^{-2 \ri L}) \big) &= e^{\ri L} \bm \nabla_{\a\b} (\cZ e^{-2 \ri L}) \cF_z{}^k_\g 
- 4 \ri \eps_{\g (\a} \bm \nabla_{\b) \ad} \big( e^{- \ri L} \bar{\bm \nabla}^{\ad k} (\bar{\cZ} e^{2 \ri L}) \big)  \ \non\\
&\quad+ \ri e^{- \ri L} \eps_{\g (\a} [\bm \nabla_{\b)}^j, \bar{\bm \nabla}_{\ad j}] L \bar{\bm \nabla}^{\ad k} (\bar{\cZ} e^{2 \ri L}) \ , \non\\
\D \cF_{\a\b} =& - \ri \D L \cF_{\a\b} - \hf e^{- \ri L} \bm \nabla_{(\a}^k \cZ \D\bm \nabla_{\b ) k} L - \frac{1}{4} \cZ e^{- \ri L} \D \bm \nabla_{\a\b} L \non\\
&+ \ri \cZ e^{-\ri L} \D \bm \nabla_{(\a}^k L \bm \nabla_{\b) k} L - e^{\ri L} W_{\a\b} \bar{\cZ} \D L \ .
\end{align}

An interesting observation, discussed in subsection \ref{VTM}, was that the second constraint \eqref{VTConstraint2R} can be motivated from the 
first constraint \eqref{VTConstraint1} by a consistency requirement. One would expect a similar situation for the variant VT multiplet. Starting with 
the condition that $L$ is independent of the harmonics
\be \bm \nabla^{--} L = 0 \ ,
\ee
we apply successive gauged covariant derivatives. Their (anti-)commutation relations lead to the consistency requirement
\begin{align}
0 &= {\bm\nabla}^{\a+} {\bm\nabla}_\a^{+} \bar{{\bm\nabla}}_\ad^{+} \bar{{\bm\nabla}}^{\ad +} \bm \nabla^{--} L \non\\
&= \bm \nabla^{--} {\bm\nabla}^{\a+} {\bm\nabla}_\a^{+} \bar{{\bm\nabla}}_\ad^{+} \bar{{\bm\nabla}}^{\ad +} L 
+ 8 \ri {\bm\nabla}^{\a\ad} {\bm\nabla}_\a^{+} \bar{{\bm\nabla}}^{+}_\ad L 
	\eol & \quad
- 2 {\bm\nabla}^{\a-} {\bm\nabla}_\a^{+} \bar{{\bm\nabla}}_\ad^{+} \bar{{\bm\nabla}}^{\ad+} L
- 2 \bar{{\bm\nabla}}_\ad^{-} \bar{{\bm\nabla}}^{\ad+} {\bm\nabla}^{\a+} {\bm\nabla}_\a^{+} L \non\\
& \quad + 2 \D \Big( e^{\ri L} ({\bm\nabla}^{+})^2 (\cZ e^{- 2 \ri L}) + e^{-\ri L} (\bar{\bm\nabla}^{+})^2 (\bar{\cZ} e^{2 \ri L}) \Big) \non\\
& - 2 \ri \D L \Big( e^{\ri L} ({\bm\nabla}^{+})^2 (\cZ e^{- 2 \ri L}) + e^{-\ri L} (\bar{\bm\nabla}^{+})^2 (\bar{\cZ} e^{2 \ri L}) \Big) \ ,
\end{align}
which is clearly satisfied by the second constraint in \eqref{CCVT2}. In the above we have made use of the following identity:
\begin{align} \bm \nabla_\a^i \bm \nabla_\b^j =& \hf (\eps_{\a\b} \bm \nabla^{ij} - \eps^{ij} \bm \nabla_{\a\b}) + \ri \eps^{ij} \eps_{\a\b} \bar{\cZ} e^{\ri L} \D 
+ \eps^{ij} \eps_{\a\b} \bar{W}_{\gd\dd} \bar{M}^{\gd \dd} \non\\
&- \frac{1}{4} \eps^{ij} \eps_{\a\b} \bar{\bm \nabla}_{\gd k} \bar{W}^{\gd\dd} \bar{S}^k_\dd - \frac{1}{4} \eps^{ij} \eps_{\a\b} \bm \nabla_{\g\dd} \bar{W}^\dd{}_\gd K^{\g\gd} \ .
\end{align}

The constraints imposed still do not define an irreducible multiplet, and an additional constraint is required. This issue will be addressed in the next subsection.

%%%%%%%%%%%%%%%%%%%%%%%%%%%%%%%%%%%%%%%%%%%%%%%%%%%%%%

\subsection{Gauge two-form}

Building on the insight gained from the standard VT multiplet we will impose an additional constraint with the use of a gauge two-form. We 
introduce the three-form field strength for the variant VT multiplet
\be H = \bm \nabla B - \hf \eta_{\hat{I}\hat{J}} V^{\hat{I}} \rd V^{\hat{J}} \ , \label{VV3form}
\ee
where $B$ is the gauge two-form and $F^{\hat{I}} = (F, F^I)$ is a set of vector field strengths with corresponding gauge one-forms $V^{\hat{I}} = (V, V^I)$.\footnote{Here the field strength $F$ 
and its gauge potential $V$ corresponds to the vector field $\cZ$.} The three-form 
field strength $H$ then satisfies the Bianchi identity
\be
\bm \nabla_{[\cA} H_{\cB\cC\cD\}} - \frac{3}{2} T_{[\cA\cB}{}^\cE H_{|\cE|\cC\cD \}} 
+ \frac{3}{4} \eta_{\hat{I}\hat{J}} F^{\hat{I}}_{[\cA\cB} F^{\hat{J}}_{\cC\cD\}} = 0 \ . \label{HBI2}
\ee
Note that due to the Bianchi identities satisfied by $\cF_{AB} = T_{AB}{}^z$, introducing terms like $F \cF$  or $\cF \cF$ on the right hand side of \eqref{HBI2} 
(and their corresponding terms in \eqref{VV3form}) only shifts the $H_{z AB}$ components and does not impose additional constraints. We now impose similar 
constraints to that for the standard VT multiplet
\begin{align} H_{\underline{\a} \underline{\b} \underline{\g}} &= H_{\underline{\ad} \underline{\bd} \underline{\gd}} 
= H_{\underline{\a} \underline{\b} \underline{\gd}} = H_{\underline{\a} \underline{\bd} \underline{\gd}} =  0 \ , \non\\
H_{a \underline{\b} \underline{\g}} &= H_{a \underline{\bd} \underline{\gd}} = 0 \ , \quad H_a{}_\b^j{}^\gd_k = - 2 \ri \d^j_k (\s_a)_\b{}^\gd \tilde{H} \ ,
\end{align}
where $\tilde{H}$ is only required to be a real superfield. The Bianchi identities then lead to the additional constraint
\begin{align}
e^{- \ri L}\bm \nabla^{ij} (\cZ K e^{2 \ri L}) + e^{\ri L} \bar{\bm \nabla}^{ij} (\bar{\cZ} \bar{K} e^{- 2 \ri L}) = 0 \ , \quad K \equiv \eta_{11} + \eta_{1I} \frac{\cW^I}{\cZ}
+ \eta_{IJ} \frac{\cW^I \cW^J}{\cZ^2} \ ,
\end{align}
and constrain the lower mass-dimension components to be
\begin{align}
H_z{}_\a^i{}_\b^j &= - \ri \eps_{\a\b} \eps^{ij} \bar{\cZ} \bar{K} e^{- \ri L} \ , \quad H_z{}^\ad_i{}^\bd_j = - \ri \eps^{\ad\bd} \eps_{ij} \cZ K e^{\ri L} \ , \non\\
H_{z \underline{\a} \underline{\bd}} &= 0 \ , \quad H_a{}_\b^j{}^\gd_k = - 2 \ri \d^j_k (\s_a)_\b{}^\gd \tilde{H} \ ,
\end{align}
where
\begin{align}
\tilde{H} =&- \frac{1}{8} \cZ \bar{\cZ} \big( 2 \eta_{11} (\cos(2L) + 1) + \eta_{1I} (1+ e^{2 \ri L}) \frac{\cW^I}{\cZ} + \eta_{1I} (1+ e^{-2 \ri L}) \frac{\bar{\cW}^I}{\bar{\cZ}} \non\\
&+ \eta_{IJ} (e^{2 \ri L} \frac{\cW^I \cW^J}{\cZ^2} + e^{-2 \ri L} \frac{\bar{\cW}^I \bar{\cW}^J}{\bar{\cZ}^2} + 2 \frac{\cW^I \bar{\cW}^J}{\cZ \bar{\cZ}})\big) \ .
\end{align}
The higher mass-dimension components are then found to be
\begin{align}
H_{a z}{}_\g^k &= \frac{1}{4} (\s_a)_\g{}^\ad e^{\ri L} \bar{\bm \nabla}_\ad^k ( \bar{\cZ} \bar{K} e^{- 2 \ri L} ) \ , \quad H_{a z}{}^\gd_k = \frac{1}{4} (\s_a)_\a{}^\gd e^{- \ri L} \bm \nabla^\a_k (\cZ K e^{2 \ri L}) \ , \non\\
H_{ab}{}_\g^k &= 2 (\s_{ab})_\g{}^\a \bm \nabla_\a^k \tilde{H} \ , \quad H_{ab}{}^\gd_k = 2 (\tilde{\s}_{ab})^\gd{}_\ad \bar{\bm \nabla}^\ad_k \tilde{H} \ , \non\\
H_{ab z} &= \frac{\ri}{16} (\s_{ab})^{\a\b} e^{- \ri L} (\bm \nabla_{\a\b} (\cZ K e^{2 \ri L}) + 4 W_{\a\b} \bar{\cZ} \bar{K}) + \HC \ , \non\\
H_{abc} &=  \frac{1}{16} \eps_{abcd} (\s^d)^{\a\ad} \Big(\frac{\ri}{2} \cZ \bar{\cZ} (K e^{2 \ri L} - \bar{K} e^{-2 \ri L}) [\bm \nabla_\a^k , \bar{\bm \nabla}_{\ad k}] L 
+ 2 \eta_{11} \bm \nabla_\a^k \cZ \bar{\bm \nabla}_{\ad k} \bar{\cZ} \non\\
&+ \eta_{1I} \bm \nabla_\a^k \cW^I \bar{\bm \nabla}_{\ad k} \bar{\cZ} + \eta_{1I} \bm \nabla_\a^k \cZ \bar{\bm \nabla}_{\ad k} \bar{\cW}^I
+ 2 \eta_{IJ} \bm \nabla_\a^k \cW^I \bar{\bm \nabla}_{\ad k} \bar{\cW}^J \non\\
&- e^{2 \ri L} \bar{\bm \nabla}_{\ad k} \bar{\cZ} \bm \nabla_\a^k (\cZ K) - 2 \ri \cZ K e^{2 \ri L} \bar{\bm \nabla}_{\ad k} \bar{\cZ} \bm \nabla_\a^k L 
- 2 \ri \bar{\cZ} e^{2 \ri L} \bar{\bm \nabla}_{\ad k} L \bm \nabla_\a^k (\cZ K) \non\\
&- 4 \cZ \bar{\cZ} (K e^{2 \ri L} + \bar{K} e^{- 2 \ri L}) \bm \nabla_\a^k L \bar{\bm \nabla}_{\ad k} L + e^{ - 2 \ri L} \bm \nabla_\a^k \cZ \bar{\bm \nabla}_{\ad k} (\bar{\cZ} \bar{K}) \non\\
&- 2 \ri e^{- 2 \ri L} \bar{\cZ} \bar{K} \bm \nabla_\a^k \cZ \bar{\bm \nabla}_{\ad k} L - 2 \ri \cZ e^{-2 \ri L} \bm \nabla_\a^k L \bar{\bm \nabla}_{\ad k} (\bar{\cZ} \bar{K}) \Big) \non\\
& + \frac{\ri}{8} \eps_{abcd} \Big( 2 \eta_{11} \bar{\cZ} \bm \nabla^d \cZ - 2 \eta_{11} \cZ \bm \nabla^d \bar{\cZ} - \eta_{1I} \cW^I \bm \nabla^d \bar{\cZ}  
+ \eta_{1I} \bar{\cZ} \bm \nabla^d \cW^I \non\\
&- \eta_{1I} \cZ \bm \nabla^d \bar{\cW}^I + \eta_{1I} \bar{\cW}^I \bm \nabla^d \cZ - 2 \eta_{IJ} \cW^I \bm \nabla^d \bar{\cW}^J + 2 \eta_{IJ} \bar{\cW}^I \bm \nabla^d \cW^J \non\\
&+ \cZ K \bm \nabla^d \bar{\cZ} e^{2 \ri L} - \bar{\cZ} \bm \nabla^d(\cZ K) e^{2 \ri L} + \cZ \bm \nabla^d (\bar{\cZ} \bar{K}) e^{- 2 \ri L} - \bar{\cZ} \bar{K} \bm \nabla^d \cZ e^{- 2 \ri L} \Big) \ .
\end{align}

The last few Bianchi identities
\begin{align}
\D H_{ab \underline{\g}} &= \bm \nabla_{\underline{\d}} H_{ab z} - 2 \bm \nabla_{[a} H_{b] z \underline{\g}}  - T_{ab}{}^{\underline{\d}} H_{z \underline{\g} \underline{\d}} 
- 2 T_{\underline{\g} [a}{}^{\underline{\dd}} H_{b] z \underline{\dd}} \non\\
&+ \cF_{az} H_{b z \underline{\g}} + \cF_{bz} H_{a z \underline{\g}} + \cF_{z \underline{\g}} H_{a b z} \ , \non\\
2 \bm \nabla_{\underline{\g}} H_{abc} &= 6 \bm \nabla_{[a} H_{bc] \underline{\g}} + 6 \cF_{[ab} H_{c] z \underline{\g}} + 6 T_{[ab}{}^{\underline{\d}} H_{c] \underline{\g} \underline{\d}} 
+ 6 T_{[ab}{}^{\underline{\dd}} H_{c] \underline{\g} \underline{\dd}} + 6 T_{\underline{\g} [a}{}^{\underline{\dd}} H_{bc] \underline{\dd}} \non\\
&+ 3 \eta_{\hat{I}\hat{J}} F^{\hat{I}}_{[ab} F^{\hat{J}}_{c] \underline{\g}} - 3 \eta_{\hat{I}\hat{J}} F^{\hat{I}}_{\underline{\g} [a} F^{\hat{J}}_{bc]} \ , \non\\
\D H_{abc} &= 3 \bm \nabla_{[a} H_{bc] z} + 3 T_{[ab}{}^{\underline{\g}} H_{c] z \underline{\g}} + 3 T_{[ab}{}^{\underline{\gd}} H_{c] z \underline{\gd}} - 3 \cF_{z [a} H_{bc] z} \ , \non\\
 \bm \nabla_{[a} H_{bcd]} &= \frac{3}{2} T_{[ab}{}^{\underline{\g}} H_{c d] \underline{\g}} + \frac{3}{2} T_{[ab}{}^{\underline{\gd}} H_{c d] \underline{\gd}} 
 + \frac{3}{2} \cF_{[ab} H_{c d] z} - \frac{3}{4} \eta_{\hat{I}\hat{J}} F^{\hat{I}}_{[ab} F^{\hat{J}}_{cd]} \ .
\end{align}
can be shown to be identically satisfied.

This completes our construction of the variant VT multiplet and its Chern-Simons couplings to vector multiplets in $\cN = 2$ supergravity.

%%%%%%%%%%%%%%%%%%%%%%%%%%%%%%%%%%%%%%%%%%%%%%%%%%%%%%
%%%%%%%%%%%%%%%%%%%%%%%%%%%%%%%%%%%%%%%%%%%%%%%%%%%%%%

\section{Discussion} \label{Conclusion}

In this paper, we have obtained several results. Firstly, we have constructed a superform formulation for the general VT multiplet system given in
\cite{Claus3} (and later lifted to superspace \cite{BN}). A remarkable feature of this formulation is that 
the constraints of \cite{BN} were shown to be reproduced from a rather simple looking superform structure and constraints on field strengths. 
Furthermore the first two constraints of \cite{BN}, equations \eqref{VTConstraint1} and \eqref{VTConstraint2R}, were shown to follow from the one-form geometry with the 
additional constraint \eqref{Constraint3} being related to the existence of the two-form coupled to the one-form with the use of Chern-Simons couplings. These results 
highlight the geometric origin of the complex expressions in \cite{Claus3}.

Our second main result is the new procedure to gauge the central charge in conformal supergravity 
-- namely, by making use of a variant VT multiplet, which has as its basic property a gauge potential with nontrivial action under the central charge. 
Its one-form geometry generalizes that of the vector multiplet. An additional constraint following from a two-form formulation is used to constrain the number 
of component fields to that of a VT multiplet. The constraints can be written in a neat form and include the presence of Chern-Simons type couplings to vector multiplets. Furthermore, 
the variant VT multiplet in the rigid supersymmetric limit and with $\eta_{1I} = \eta_{IJ} = 0$ reduces to the results found in \cite{Theis1, Theis2}.\footnote{The constraints of \cite{Theis2} 
differ by a constant shift in $L$.} Moreover it is conceivable that the new procedure of gauging the central charge will lead to new couplings of multiplets to 
supergravity.\footnote{This issue will be further discussed in \cite{BKN}.}

One of the advantages of the superform formulation is that the component one-form and two-form are built into the theory {\it ab initio}. It follows that the supersymmetry
transformations close on the component fields, which appear in the component projection of the superforms. Making use of the component reduction methods of \cite{BN} one can then 
reproduce the component results of \cite{Claus3} straightforwardly via component reduction. As a result the superform formulation 
proves to be a powerful tool in describing multiplets in supergravity.

In order to describe the dynamics of the variant VT multiplet we require an action.  
The action for the VT multiplet of \cite{Claus3} may be constructed with the use of a composite linear multiplet\footnote{It is known how to do this in the case of 
harmonic superspace \cite{KuzenkoTheisen}.}, worked out in terms of superfields in \cite{KN, BN}. For the case of the variant VT multiplet, it is already known that 
in the flat case the linear multiplet must be modified \cite{Theis2}. Motivated by the flat case,
one expects that in supergravity we need a real isospinor superfield, $\cL^{ij}$, satisfying a modified constraint 
\be \bm \nabla_\a^{(i} (e^{\ri L} \cL^{jk)}) = 0 \ . \label{ModLinearity}
\ee
The reason for this modification will be discussed in \cite{BKN}.

In order to obtain the action corresponding to the variant VT multiplet we need to find a composite $\cL^{ij}$ built from the basic superfields. Considering the 
constraints we can immediately write down a candidate for the Lagrangian
\be \cL^{ij} = \frac{\ri}{4} e^{-\ri L} \bm \nabla^{ij} (\cZ K e^{2 \ri L}) = - \frac{\ri}{4} e^{\ri L} \bar{\bm \nabla}^{ij} (\bar{\cZ } \bar{K} e^{- 2 \ri L}) \ .
\ee
However in the flat case (and with $\eta_{1I} = \eta_{IJ} = 0 $) this can be shown to correspond to the total derivative Lagrangian in \cite{Theis2}. It was noted in \cite{Theis2} that 
the Lagrangian found from considering an ansatz possessed a particular symmetry. This is to be expected since the constraints remain invariant 
under the shift $L \rightarrow L + 2 \pi$ and so we expect the action to also possess such a symmetry. Thus a shift by $2 \pi$ can only shift the Lagrangian 
by a total derivative.\footnote{The analogous case for the standard VT multiplet was briefly discussed in subsection \ref{VTM}.} This leads one to consider a candidate for the Lagrangian of the form
\be \cL^{ij} = \frac{\ri}{2} e^{-\ri L} \bm \nabla^{ij} (\cZ K L e^{2 \ri L}) + \cdots + \HC \ ,
\ee
where $\cdots$ represents terms that are invariant under the shift. Then one can show that the `linearity' condition \eqref{ModLinearity} fixes the remaining terms
\begin{align}
\cL^{ij} &= \frac{\ri}{2} e^{-\ri L} \bm \nabla^{ij} (\cZ K L e^{2 \ri L}) - e^{- \ri L} \bar{\cZ} \bar{K} \bar{\bm \nabla}^i L \bar{\bm \nabla}^j L - \frac{1}{4} e^{- \ri L} \bar{\bm \nabla}^{ij}(\bar{\cZ} \bar{K}) + \HC
\end{align}
This Lagrangian is a new result and corresponds to the variant VT multiplet with Chern-Simons type couplings in supergravity, reducing in the flat case (and with $\eta_{1I} = \eta_{IJ} = 0$) 
to the result found in \cite{Theis2}. The corresponding action principle will be discussed in a separate paper \cite{BKN}.

%%%%%%%%%%%%%%%%%%%%%%%%%%%%%%%%%%%%%%%%%%%%%%%%%%%%%%
%%%%%%%%%%%%%%%%%%%%%%%%%%%%%%%%%%%%%%%%%%%%%%%%%%%%%%

\noindent
{\bf Acknowledgements:}\\
I would like to thank Sergei Kuzenko for ongoing supervision and providing valuable feedback on the manuscript. I would also like to thank Daniel Butter for
numerous stimulating and helpful discussions as well as reading the manuscript; and Ulrich Theis for correspondence. This work is supported by an Australian Postgraduate Award.

%%%%%%%%%%%%%%%%%%%%%%%%%%%%%%%%%%%%%%%%%%%%%%%%%%%%%%
%%%%%%%%%%%%%%%%%%%%%%%%%%%%%%%%%%%%%%%%%%%%%%%%%%%%%%

\appendix

%%%%%%%%%%%%%%%%%%%%%%%%%%%%%%%%%%%%%%%%%%%%%%%%%%%%%%
%%%%%%%%%%%%%%%%%%%%%%%%%%%%%%%%%%%%%%%%%%%%%%%%%%%%%%

\section{Conformal superspace} \label{conformalSpace}

This appendix contains a brief summary of conformal superspace of \cite{Butter4D}.\footnote{The conventions here differ; see however \cite{BN}.}
 Consider a curved 4D $\cN = 2$ superspace $\cM^{4|8}$ parametrized by local bosonic $(x)$ 
and fermionic $(\q, \bar{\q})$ coordinates 
$z^M = (x^m, \ \q^\mu_\imath, \ \bar{\q}_{\dot{\mu}}^\imath)$, where
$m = 0, 1, \cdots, 3,$ $\mu = 1, 2$, $\dot{\mu} = 1, 2$ and $\imath = \1, \2$. The
Grassmann variables $\q^\mu_\imath$ and $\bar{\q}_{\dot{\mu}}^\imath$ are related
to each other by complex conjugation: $\overline{\q^\mu_\imath} = \bar{\q}^{\dot{\mu} \imath}$.
The structure group is chosen to be $\rm SU(2, 2|2)$ and the covariant derivatives
$\nabla_A = (\nabla_a, \nabla_\a^i , \bar{\nabla}^\ad_i)$ have the form
\begin{align} \nabla_A &= E_A + \hf \Omega_A{}^{ab} M_{ab} + \Phi_A{}^{ij} J_{ij} + \ri \Phi_A Y 
+ B_A \mathbb{D} + \frak{F}_{A}{}^B K_B \non\\
&= E_A + \Omega_A{}^{\b\g} M_{\b\g} + \bar{\Omega}_A{}^{\bd\gd} \bar{M}_{\bd\gd} + \Phi_A{}^{ij} J_{ij} + \ri \Phi_A Y 
+ B_A \mathbb{D} + \frak{F}_{A}{}^B K_B \ .
\end{align}
Here $E_A = E_A{}^M(z) \partial_M$ is the supervielbein, with $\partial_M = \partial/\partial z^M$,
$J_{kl} = J_{lk}$ are generators of the group $\rm SU(2)_R$,
$M_{ab}$ are the Lorentz generators, $Y$ is the generator of the chiral rotation
group $\rm U(1)_R$, and $K^A = (K^a, S^\a_i, \bar{S}_\ad^i)$ are the special
superconformal generators. The one-forms $\Omega_A{}^{bc}$, $\Phi_A{}^{kl}$, $\Phi_A$, $B_A$ and
$\frak{F}_A{}^B$ are the corresponding connections.

The generators act on the covariant derivatives as
\begin{align}
[M_{ab}, \nabla_c ] &= 2 \eta_{c [a} \nabla_{b]}~, \quad [M_{ab}, \nabla_\a^i] = (\s_{ab})_\a{}^\b \nabla_\b^i ~, \quad
[M_{ab}, \bar\nabla^\ad_i] = (\tilde{\s}_{ab})^\ad{}_\bd \bar\nabla^\bd_i~, \non\\
[J_{ij}, \nabla_\a^k] &= - \d^k_{(i} \nabla_{\a j)} ~,\quad
[J_{ij}, \bar\nabla^\ad_k] = - \ve_{k (i} \bar\nabla^{\ad}_{j)}~, \non \\
[Y, \nabla_\a^i] &= \nabla_\a^i ~,\quad [Y, \bar\nabla^\ad_i] = - \bar\nabla^\ad_i~,  \non \\
[\mathbb{D}, \nabla_a] &= \nabla_a ~, \quad
[\mathbb{D}, \nabla_\a^i] = \hf \nabla_\a^i ~, \quad
[\mathbb{D}, \bar\nabla^\ad_i] = \hf \bar\nabla^\ad_i ~ .
\end{align}

Finally, the algebra of $K^A$ with $\nabla_B$ is given by
\begin{align}
[K^a, \nabla_b] &= 2 \delta^a_b \mathbb{D} + 2 M^{a}{}_b ~,\non \\
\{ S^\a_i , \nabla_\b^j \} &= 2 \d^j_i \d^\a_\b \mathbb{D} - 4 \d^j_i M^\a{}_\b 
- \d^j_i \d^\a_\b Y + 4 \d^\a_\b J_i{}^j ~,\non \\
\{ \bar{S}^i_\ad , \bar{\nabla}^\bd_j \} &= 2 \d^i_j \d^\bd_\ad \mathbb{D} 
+ 4 \d^i_j \bar{M}_\ad{}^\bd + \d^i_j \d_\ad^\bd Y - 4 \d_\ad^\bd J^i{}_j ~,\non \\
[K^a, \nabla_\b^j] &= -\ri (\s^a)_\b{}^\bd \bar{S}_\bd^j \ , \quad [K^a, \bar{\nabla}^\bd_j] = 
-\ri ({\s}^a)^\bd{}_\b S^\b_j ~, \non \\
[S^\a_i , \nabla_b] &= \ri (\s_b)^\a{}_\bd \bar{\nabla}^\bd_i \ , \quad [\bar{S}^i_\ad , \nabla_b] = 
\ri ({\s}_b)_\ad{}^\b \nabla_\b^i \ ,
\end{align}
where all other (anti-)commutations vanish.

The covariant derivatives obey the (anti-)commutation relations:
\begin{subequations}\label{CSGAlgebra}
\begin{align}
\{ \nabla_\a^i , \nabla_\b^j \} &= 2 \ve^{ij} \ve_{\a\b} \bar{W}_{\gd\dd} \bar{M}^{\gd\dd} + \hf \ve^{ij} \ve_{\a\b} \bar{\nabla}_{\gd k} \bar{W}^{\gd\dd} \bar{S}^k_\dd 
- \hf \ve^{ij} \ve_{\a\b} \nabla_{\g\dd} \bar{W}^\dd{}_\gd K^{\g \gd}~, \\
\{ \bar{\nabla}^\ad_i , \bar{\nabla}^\bd_j \} &= - 2 \ve_{ij} \ve^{\ad\bd} W^{\g\d} M_{\g\d} + \frac{1}{2} \ve_{ij} \ve^{\ad\bd} \nabla^{\g k} W_{\g\d} S^\d_k 
- \frac{1}{2} \ve_{ij} \ve^{\ad\bd} \nabla^{\g\gd} W_{\g}{}^\d K_{\d \gd}~, \\
\{ \nabla_\a^i , \bar{\nabla}^\bd_j \} &= - 2 \ri \d_j^i \nabla_\a{}^\bd~, \\
[\nabla_{\a\ad} , \nabla_\b^i ] &= - \ri \ve_{\a\b} \bar{W}_{\ad\bd} \bar{\nabla}^{\bd i} - \frac{\ri}{2} \ve_{\a\b} \bar{\nabla}^{\bd i} \bar{W}_{\ad\bd} \mathbb{D} 
- \frac{\ri}{4} \ve_{\a\b} \bar{\nabla}^{\bd i} \bar{W}_{\ad\bd} Y + \ri \ve_{\a\b} \bar{\nabla}^\bd_j \bar{W}_{\ad\bd} J^{ij}
	\eol & \quad
	- \ri \ve_{\a\b} \bar{\nabla}_\bd^i \bar{W}_{\gd\ad} \bar{M}^{\bd \gd} - \frac{\ri}{4} \ve_{\a\b} \bar{\nabla}_\ad^i \bar{\nabla}^\bd_k \bar{W}_{\bd\gd} \bar{S}^{\gd k} 
	+ \frac{1}{2} \ve_{\a\b} \nabla^{\g \bd} \bar{W}_{\ad\bd} S^i_\g
	\eol & \quad
	+ \frac{\ri}{4} \ve_{\a\b} \bar{\nabla}_\ad^i \nabla^\g{}_\gd \bar{W}^{\gd \bd} K_{\g \bd}~, \\
[ \nabla_{\a\ad} , \bar{\nabla}^\bd_i ] &=  \ri \d^\bd_\ad W_{\a\b} \nabla^{\b}_i + \frac{\ri}{2} \d^\bd_\ad \nabla^{\b}_i W_{\a\b} \mathbb{D} 
- \frac{\ri}{4} \d^\bd_\ad \nabla^{\b}_i W_{\a\b} Y + \ri \d^\bd_\ad \nabla^{\b j} W_{\a\b} J_{ij}
	\eol & \quad
	+ \ri \d^\bd_\ad \nabla^{\b}_i W^\g{}_\a M_{\b\g} + \frac{\ri}{4} \d^\bd_\ad \nabla_{\a i} \nabla^{\b j} W_\b{}^\g S_{\g j} - \hf \d^\bd_\ad \nabla^\b{}_\gd W_{\a\b} \bar{S}^{\gd}_i
	\eol & \quad
	+ \frac{\ri}{4} \d^\bd_\ad \nabla_{\a i} \nabla^\g{}_\gd W_{\b\g} K^{\b\gd} ~.
\end{align}
\end{subequations}
The complex superfield $W_{\a\b} = W_{\b\a}$ and its complex conjugate
${\bar{W}}_{\ad \bd} := \overline{W_{\a\b}}$ are superconformally primary,
$K_A W_{\a\b} = 0$, and obey the additional constraints
\begin{align}
\bar{\nabla}^\ad_i W_{\b\g} = 0~,\qquad
\nabla_{\a\b} W^{\a\b} &= \bar{\nabla}^{\ad\bd} \bar{W}_{\ad\bd} ~,
\end{align}
where
\begin{align}
\nabla_{\a\b} := \nabla_{(\a}^k \nabla_{\b) k} \ , \quad \bar{\nabla}^{\ad\bd} := \nabla^{(\ad}_k \nabla^{\bd) k} \ .
\end{align}

As an easy lookup a list of the non-vanishing torsion components is given below:
\begin{align}
T_\a^i{}^\bd_j{}^a &= - 2 \ri \d^i_j (\s^a)_\a{}^\bd \ , \non\\
T_a{}^j_\b{}^k_\gd &= - \frac{\ri}{2} \eps^{jk} (\s_a)_\b{}^\bd \bar{W}_{\bd\gd} \ , \quad T_a{}^\bd_j{}^\g_k = - \frac{\ri}{2} \eps_{jk} (\s_a)_\b{}^\bd W^{\b\g} \ , \non\\
T_{ab}{}^\g_k &= \frac{1}{4} (\s_{ab})^{\a\b} \nabla^\g_k W_{\a\b} \ ,  \quad T_{ab}{}_\gd^k = \frac{1}{4} (\tilde{\s}_{ab})^{\ad\bd} \bar{\nabla}_\gd^k \bar{W}_{\ad\bd} \ .
\end{align}

%%%%%%%%%%%%%%%%%%%%%%%%%%%%%%%%%%%%%%%%%%%%%%%%%%%%%%
%%%%%%%%%%%%%%%%%%%%%%%%%%%%%%%%%%%%%%%%%%%%%%%%%%%%%%

\section{A note on solving Bianchi identities} \label{BINote}

When solving Bianchi identities for a superform one imposes constraints and solves the components in terms of superfields. Once all the components are found the 
remaining Bianchi identities remain as consistency checks. However it can be seen that beyond a certain point the remaining 
Bianchi identities are identically satisfied.\footnote{I am grateful to Daniel Butter for pointing out this procedure, which is similar to that used in supergravity ({\it e.g.} Dragon's Theorem).} 
To see this we start with a $p$-form, $H$, with a gauge potential $B$
\be H = \bm \nabla B \ .
\ee
This superform then satisfies the Bianchi identity
\be \bm \nabla_{[\cA_1} H_{\cA_2 \cdots \cA_{p+1} \} } - \frac{p}{2} T_{[\cA_1 \cA_2}{}^{\cE} H_{|\cE|\cA_3\cdots \cA_{p+1} \}} = 0 \ , 
\ee
where we adopt the 5D notation discussed in the the paper and we allow for the possibility that the central charge may be gauged with a vector multiplet or a variant VT multiplet.
Denoting
\be I_{\cA_1 \cdots \cA_{p+1}} := \bm \nabla_{[\cA_1} H_{\cA_2 \cdots \cA_{p+1} \} } - \frac{p}{2} T_{[\cA_1 \cA_2}{}^\cC H_{|\cC| \cA_3 \cdots \cA_{p+1} \}} \ ,
\ee
we want to check that $I = 0$. Its Bianchi identity can be rewritten as
\be \bm \nabla_{[\cA_1} I_{\cA_2 \cdots \cA_{p+2} \}} - \frac{p+1}{2} T_{[\cA_1 \cA_2}{}^\cC I_{|\cC| \cA_3 \cdots \cA_{p+2} \} } = 0 \ . \label{Iclosed}
\ee
Suppose now that all Bianchi identities for $H$ are satisfied except for those corresponding to
\be I_{a_1 \cdots a_{p-1} z \underline{\a}} \ , \quad I_{a_1 \cdots a_{p} \underline{\a}} \ , \quad I_{a_1 \cdots a_{p} z} \ , \quad I_{a_1 \cdots a_{p+1}} \ ,
\ee
{\rm i.e.} all lower mass-dimension components of $I$ vanish. Then setting $\cA_1 = a_1 \ , \cdots \ , \cA_{p-2} = a_ {p-2} \ , \cA_{p-1} = z \ , \cA_{p} = \underline{\a} \ , 
\cA_{p+1} = \underline{\bd} \ , \cA_{p+2} = \underline{\g}$ in equation
\eqref{Iclosed} gives
\be T_{(\underline{\a} \underline{\bd}}{}^b I_{\underline{\g}) b a_1 \cdots a_{p-2} z} = 0 \ ,
\ee
which due to the form of the torsion gives
\be I_{a_1 \cdots a_{p-1} z \underline{\a}} = 0 \ .
\ee
Similarly setting $\cA_1 = a_1 \ , \cdots \ , \cA_{p-1} = a_ {p-1} \ , \cA_{p} = \underline{\a} \ , \cA_{p+1} = \underline{\bd} \ , \cA_{p+2} = \underline{\g}$ leads to
\be I_{a_1 \cdots a_{p} \underline{\a}} = 0 \ .
\ee
Setting $\cA_1 = a_1 \ , \cdots \ , \cA_{p-1} = a_ {p-1} \ , \cA_{p} = z \ , \cA_{p+1} = \underline{\a} \ , \cA_{p+2} = \underline{\bd}$ gives
\be I_{a_1 \cdots a_{p} z} = 0 \ .
\ee
We can similarly deduce
\be
I_{a_1 \cdots a_{p+1}} = 0 \ .
\ee
Thus it follows that the remaining Bianchi identities are identically satisfied. A generalization to the Bianchi identities of the two-form geometry for both the VT multiplet and the variant VT multiplet
 discussed in the paper is straightforward.

\section{A 5D interpretation} \label{5DI}

The observation in subsection \ref{SGTS} suggests that it is natural to think of the central charge as an additional covariant derivative. In flat central charge 
superspace, Ref. \cite{HOW, GHH, BHO} showed that the central charge may be thought of as a derivative with respect to an additional bosonic coordinate, $z$, and thus we may interpret
the central charge as a transformation along $z$. The generalization to supergravity was discussed in \cite{AGHH}, where it was used to describe a particular VT multiplet. 
Here we discuss a similar construction for the case of a real central charge.

In addition to the superspace coordinates, $z^M = (x^m, \theta^\mu_i , \bar{\theta}_{\dot{\mu}}^i)$ we have a real bosonic coordinate, $z$,
\be \bm z^\cM = (x^m, \theta^\mu_i , \bar{\theta}_{\dot{\mu}}^i, z) \ .
\ee
We can then introduce a supervielbein, 
\be E_{\cA} = E_{\cA}{}^\cM \partial_\cM \ , \quad \partial_\cM = \frac{\partial}{\partial \bm z^\cM} \ ,
\ee
where $E_A{}^M$ is the usual vierbein and $E_z$ is added for the sector corresponding to the central charge. We then introduce the covariant derivatives
\be \bm \nabla_{\cA} = E_{\cA} + \Phi_{\cA} \ ,
\ee
where the connection $\Phi_{\cA}$ takes values in the structure group of the superconformal formulation of \cite{BN}. The connection $\Phi_{\cA}$ may also in principle 
contain phase transformations in the central charge sector \cite{Gaida}, however we will assume this is not the case. The supergravity transformations are 
then generated by local transformations of the form
\begin{align} \d_{\hat{\cK}} \bm \nabla_{\cA} &= [\hat{\cK}, \bm \nabla_{\cA}] \ , \non\\
\hat{\cK} &= \cK^\cC(z) \bm \nabla_\cC + \hf \cK^{cd}(z) M_{cd} + \cK^{kl}(z) J_{kl} + \ri \cK_Y Y + \cK_{\mathbb D} \mathbb D + \cK^A K_A \ ,
\end{align}
where the parameters satisfy the usual reality conditions.

The covariant derivative algebra is of the form
\begin{align} [\bm \nabla_\cA, \bm \nabla_\cB \} 
         &= T_{\cA\cB}{}^\cC \bm \nabla_\cC + \hf R_{\cA\cB}{}^{cd} M_{cd} + R_{\cA\cB}{}^{kl} J_{kl}
	\eol & \quad
	+ \ri R_{\cA\cB}(Y) Y + R_{\cA\cB} (\mathbb{D}) \mathbb{D} + R_{\cA\cB}{}^C K_C \ .
\end{align}

Now we define a gauge one-form, $\cV = E^\cA \cV_\cA$, with transformation law
\be \d \cV_\cA = \hat{\cK} \cV_\cA + \bm \nabla_{\cA} \G \ ,
\ee
where $\G$ is the gauge parameter associated with $\cV$. Its two-form field strength, $\cF = \hf E^\cB E^\cA \cF_{\cA \cB}$, is
\be \cF := \hat{\rd} \cV \ ,
\ee
where $\hat{\rd}$ is the 5D exterior derivative
\be \hat{\rd} = \hat{\rd} z^\cM \partial_{\cM} \ .
\ee

We may introduce a gauge two-form, $B= \hf E^\cB E^\cA B_{\cA \cB}$, with the transformation law
\be \d_{\hat{\cK}} B_{AB} = \hat{\cK} B_{AB} + \eta \G \cF_{\cA\cB} + 2 \bm \nabla_{[\cA} \L_{\cB\} } - T_{\cA\cB}{}^\cC \L_\cC \ ,
\ee
where $\L$ generates the gauge transformation of B. Its three-form field strength, $H$, is
\be H := \hat{\rd} B - \eta \cV \cF \ ,
\ee

The Bianchi identities are then
\begin{align} \hat{\rd} \cF &= 0 \ , \quad \bm \nabla_{[\cA} \cF_{\cB\cC\}} - T_{[\cA\cB}{}^D \cF_{|\cD|\cC\}} = 0 \ , \non\\
\hat{\rd} H &= - \hf \eta \cF \cF \ , \quad \bm \nabla_{[\cA} H_{\cB\cC\cD\}} - \frac{3}{2} T_{[\cA\cB}{}^\cE H_{|\cE|\cC\cD \}} + \frac{3}{4} \eta \cF_{[\cA\cB} \cF_{\cC\cD\}} = 0 \ .
\end{align}

The component expressions for the Bianchi identities above are formally the same as those derived in the paper under particular identifications. However, the superspace structure 
contains too many component fields and must be constrained via a choice of constraints. A choice of constraints may be motivated by the structure of 4D curved superspace with 
gauged central charge (for the vector and variant VT cases). We will leave the analysis of the constrained geometry for a discussion elsewhere.

\begin{footnotesize}

\end{footnotesize}

\end{document}